# Massive MIMO and Millimeter Wave for 5G Wireless HetNet: Potentials and Challenges

Tadilo Endeshaw Bogale, *Member, IEEE* and Long Bao Le, *Senior Member, IEEE*

*Abstract*— There have been active research activities worldwide in developing the next-generation 5G wireless network. The 5G network is expected to support significantly large amount of mobile data traffic and huge number of wireless connections, achieve better cost- and energy-efficiency as well as quality of service (QoS) in terms of communication delay, reliability and security. To this end, the 5G wireless network should exploit potential gains in different network dimensions including super dense and heterogeneous deployment of cells and massive antenna arrays (i.e., massive multiple input multiple output (MIMO) technologies) and utilization of higher frequencies, in particular millimeter wave (mmWave) frequencies. This article discusses potentials and challenges of the 5G heterogeneous wireless network (HetNet) which incorporates massive MIMO and mmWave technologies. We will first provide the typical requirements of the 5G wireless network. Then, the significance of massive MIMO and mmWave in engineering the future 5G HetNet is discussed in detail. Potential challenges associated with the design of such 5G HetNet are discussed. Finally, we provide some case studies, which illustrate the potential benefits of the considered technologies.

## I. INTRODUCTION

Research on next-generation 5G wireless systems, which aims to resolve several unprecedented technical requirements and challenges, has attracted growing interests from both academia and industry in the past few years. More than 5 billion devices demand wireless connections that runs voice, data, and other applications in today's wireless networks [1]. The study performed by Wireless World Research Forum (WWRF) has predicted that 7 trillion wireless devices will be served by wireless networks, both for human and machine-type communications, in 2017. Furthermore, the amount of mobile data traffic has increased dramatically over the years, which is mainly driven by the massive demand of data-hungry devices such as smart phones, tablets and broadband wireless applications such as multimedia, 3D video games, e-Health, Car2X communications [1]–[3]. This trend will continue and it is expected that the 5G wireless network will deliver about 1000 times more capacity than that of the current 4G system.

In addition, significant improvement in communications quality of service (QoS) is expected in the 5G network. In particular, real time support will be the key requirement to realize many emerging wireless applications. In fact, real time is a highly subjective term and depends on specific use case. According to [4], a service can be defined as real time when the communication response time (i.e., round-trip latency) is faster than the time constants of the application. Moreover, different use cases require different round-trip latency of interactions. For instance, the latency requirement for audio signal is around 80ms which can be supported by the current Long-term evolution (LTE) with typical round-trip latency of 25 ms [1], [5]. Although the LTE latency is fairly sufficient for most current services, it is anticipated that a sheer number of new use cases which require very small latency will arise in the future 5G network including two-way gaming, virtual and enhanced reality (e.g., networked wearable computed devices), and touch screen enabled applications (i.e., a tactile internet). Among these use cases, the tactile internet requires a more stringent latency requirement which is in the order of 1ms [4], [6]. In fact, it has been predicted that tactile internet will influence our daily lives and will transform different important socioeconomic sectors such as health care, education, smart grid and intelligent transportation.

Moreover, there is a strong need to achieve these technical requirements while improving the cost and energy efficiency of the future wireless network. In fact, the exponential growth of mobile data in recent years contradicts the flattening of the revenue of mobile operators. Thus, it is very desirable that the next-generation 5G network can realize cost efficient, which is measured in \$/bit, wireless technologies. In fact, energy related costs account for significant portion of the overall operational expenditure (OPEX) of wireless operators [7]. In particular, more than 70% of the mobile operator's electricity bill is due to the radio part of the wireless cellular network [7], [8]. Also, the Carbon dioxide ($CO_2$) contribution of the telecommunications sector to the global CO2 emission has increased rapidly over the last decade where mobile operators are among the top energy consumers. Thus, apart from spectral efficiency, energy efficiency is a crucial design objective to reduce the operation cost for mobile operators as well as to minimize the environmental impact of the wireless domain.

To resolve the aforementioned challenges, it becomes essential to adopt a network infrastructure that can efficiently integrate various disruptive wireless technologies and to enable inter-networking of existing and newly-deployed technologies. Such development should consider the emerging wireless applications and services in the short, medium, and long terms. In particular, the 5G network should enable us to realize the truly networked society with unlimited access of information for anyone, anywhere, and anytime. It should also allow to support various smart infrastructures and smart cities that are green, safe, mobile, connected, and informed [1]. The air interface spectral efficiency, available spectrum bands and the number of deployed base stations (BSs) are the key contrib-

The authors are with the Institute National de la Recherche Scientifique (INRS), Montreal, QC, Canada. E-mails: {tadilo.bogale, long.le}@emt.inrs.ca

utors to the performance of any network. Towards this end, it is undoubtedly important to realize network densification in multiple dimensions including deployment of super-dense HetNets with different types of cells, multiple radio access technologies (RATs), massive multiple input multiple output (MIMO) at BSs and/or user equipments (UEs) as well as exploitation of both microwave and mmWave frequency bands. The realization of these technologies will lead to the full-scale 5G HetNet, which poses various challenges at the architecture as well as communications and networking levels.

In this article, we first provide the concise description of technical requirements of the 5G wireless network. Comprehensive discussions on the considered 5G HetNet and its associated research challenges are then provided. After that we describe how massive MIMO and mmWave technologies can help address some of the challenges via case studies. In the following, we use the terms node and BS interchangeably.

## II. REQUIREMENTS OF 5G WIRELESS NETWORK

The 5G wireless network has not been standardized yet. The detailed and exact technical specifications of this network would only be available in the near future. However, the following technical requirements are accepted by wireless industries and academia [1]–[3].

- **Coverage and Data rate:** It is believed that the 5G should maintain connectivity anytime and anywhere with a minimum user experience data rate of 1Gb/sec [9]. In general, as the low mobility UEs channel changes much slower than those of high mobility ones, these UEs require more resources for CSI acquisition (i.e., reduced effective data rate). Therefore, the peak data rates required by high and low mobility users in the 5G network can be different. The network must also ensure a certain QoS for users traveling at very high speed (e.g., high speed trains traveling at 500 km/hr) where the existing networks cannot satisfactorily support[1].
- **Latency:** The latency requirement is usually more difficult to achieve compared to that of the data rate as it demands that the data be delivered to the destination within a given period of time. For the 5G network, the end-to-end latency requirement will be in the order of 1-5ms [2], [3], [6].
- **Connected devices:** The future 5G network is expected to incorporate massive amount of connected devices which may reach up to 100 times that of the current wireless network. The most potential use cases in this regard are wearable computing, machine type communications, wireless sensors, and internet of things [1], [7]. Importantly, these connected devices may have different requirements in terms of communication rate, delay and reliability.
- **Multiple RATs:** The 5G network would not be developed to replace current wireless networks. It is rather to advance and integrate the existing network infrastructures with the new one. In the 5G network, the existing wireless technologies including Global system for mobile communications (GSM), 3G, High Speed Packet Access (HSPA), LTE and LTE-advanced as well as WiFi will continue to evolve and be integrated into a unified system [1], [7].
- **Energy and cost efficiency:** 5G wireless technologies must be designed to achieve significantly better cost efficiency measured in bit/$ in order to address the revenue flattening concerns of mobile operators. In particular, the energy efficiency measured in bit/Joule of the 5G network may need to be reduced by a factor of 1000 compared to that achieved by current wireless technologies [1].

## III. 5G WIRELESS HETNET BASED ON MMWAVE AND MASSIVE MIMO

Network densification via massive deployment of cells of different types such as macrocells, microcells, picocells, and femto cells is a key technique to enhance the network capacity, coverage performance, and energy efficiency. This densification approach has been adopted in existing wireless cellular networks in particular 3G and 4G LTE systems, which essentially results in a multi-tier cellular HetNet [8], [10]. Wireless HetNets may also comprise remote radio heads (RRHs) and wireless relays, which can further boost the network performance. It is anticipated that relaying and multihop communications will be among central elements of the 5G wireless architecture (in contrast to the existing LTE system where multihop communications have been considered as an additional feature) [3]. In general, radio resource management for HetNets plays a crucial role in achieving the benefits of this advanced architecture. Specifically, development of resource allocation algorithm that efficiently utilizes radio resources including bandwidth, transmission power and antenna while mitigating inter-cell and inter-user interference and guarantees acceptable QoS for active users is one of the most critical issues. In addition, design and deployment of reliable backhaul networks that enable efficient resource management and coordination are also very important.

It is believed that massive MIMO and mmWave technologies provide vital means to resolve many technical challenges of the future 5G HetNet and they can seamlessly be integrated with the current networks and access technologies. The deployment of the massive number of antennas at the transmitter and/or receiver (massive MIMO) can significantly enhance the spectral and energy efficiency of the wireless network [11], [12]. In a rich scattering environment, these performance gains can be achieved with simple beamforming strategies such as maximum ratio transmission (MRT) or zero forcing (ZF) [11]. Moreover, most today's wireless systems operate at microwave frequencies below 6GHz. The sheer capacity requirement of the next-generation wireless network would inevitably demand us to exploit the frequency bands above 6GHz where the mmWave frequency ranging 30GHz-300GHz can offer huge spectrum, which is still under-utilized [1], [2], [13]. Most importantly, as the mmWaves have extremely short wavelength, it becomes possible to pack a large number of antenna elements in a small area, which consequently help realize massive MIMO at both the BSs and UEs.

---

[1]The 4G network can support the mobility of up to 250 km/hr.

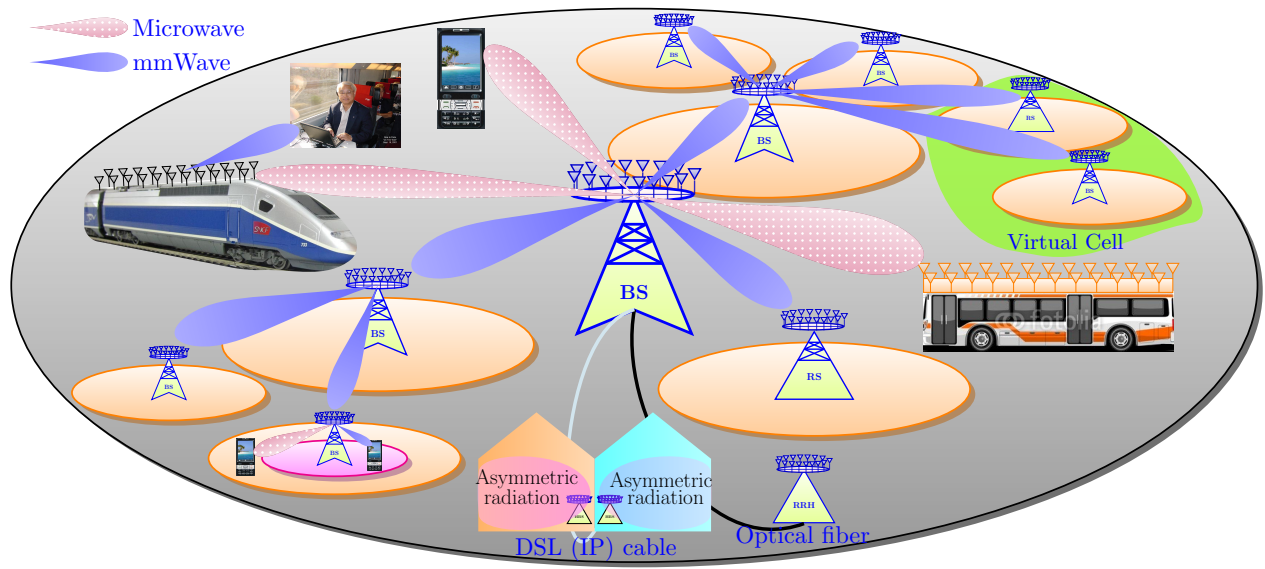

Fig. 1. Potential 5G HetNet network architecture incorporating massive MIMO and mmWave.

In particular, mmWave frequencies can be used for outdoor point-to-point backhaul links or for supporting indoor high-speed wireless applications (e.g., high-resolution multimedia streaming). In fact, mmWave technologies have already been standardized for short-range services in IEEE 802.11ad. However, these frequencies have not been well explored for cellular applications. Some potential reasons are the high propagation loss, penetration loss, rain fading and these frequencies are easily absorbed or scattered by gases [14]. The massive deployment of small cells such as pico and femto cells in the future 5G HetNet renders the short-range mmWave technologies very useful. Therefore, the mmWave frequencies can be considered as one of the potential technologies to meet the requirements of the 5G network.

There are many possibilities to enable the 5G wireless HetNet incorporating massive MIMO and mmWave technologies. One such 5G network architecture is shown in Fig. 1 where we demonstrate how massive MIMO and mmWave technologies can be used in different parts and communication purposes. The architecture of Fig. 1 employs both millimeter wave and microwave frequencies. To determine the operating frequency bands of different communications in the architecture of Fig. 1, several factors may need to be considered such as the regulatory issues, application, channel, and path loss characteristics of various frequency bands. In general, path loss increases as the carrier frequency increases. This observation leads to the utilization of microwave frequencies for long-range outdoor communications.

In mmWave frequency bands, different frequencies have distinct behaviors. For example, naturally oxygen molecule ($O_2$) absorbs electromagnetic energy at 60GHz to a much higher degree than in the regions $(30-160)$GHz in general. This absorption weakens (attenuates) 60GHz signals over distance significantly; thus, the signals cannot reach far away users. This makes the 60GHz suitable for high data-rate and secure indoor communications. Hence, selection of operating frequency depends on several factors such as application, different absorptions and blockages. Given these factors, however, there is a general consensus that mmWave frequency bands (30-300)GHz can be useful for backhaul links, indoor, short range and line of sight (LOS) communications.

Generally, the deployment of multiple antennas at the transmitter and/or receiver improves the overall performance of a wireless system. This performance improvement is achieved when the channel coefficients corresponding to different transmit-receive antennas experience independent fading. For a given carrier frequency, such independent fading channel is exhibited when the distance between two antennas is at least $0.5\lambda$, where $\lambda$ is the wavelength [12], [15]. Thus, for fixed spatial dimension, the number of deployed antennas increases as the carrier frequency increases which consequently allows to pack a large number of antennas at mmWave frequencies. Moreover, the deployment of massive MIMO can be realized at different transportation systems such as trains and buses even at microwave frequency bands since sufficient space is available to do so (i.e., as in Fig. 1).

In recent years, three dimensional (3D) and full dimensional (FD) MIMO techniques have been promoted to increase the overall network efficiency as they allow cellular systems to support a large number of UEs using multiuser MIMO techniques. Thus, the massive MIMO and mmWave systems of the considered architecture can also be designed to be either 3D or FD. On the other hand, this architecture can support coordinated multipoint (CoMP) transmission where BSs are coordinated using either fiber or wireless backhauls. In addition, Fig. 1 incorporates a cell virtualization concept where the virtual cell can be defined either by the network (network centric) or the users (user-centric), and can also be integrated as part of the cloud radio access network [16], [17].

## IV. RESEARCH CHALLENGES

The 5G network presents significantly enhanced requirements compared to those in existing wireless networks. While it is expected that the wireless HetNet incorporating massive MIMO and mmWave technologies enables us to meet the technical requirements of the 5G network, there are various challenges one has to tackle. In the following, some key challenges and the potential solution approaches are discussed.

### A. Network Planning and Traffic Management

Network planning considering the future mobile data traffic demand for the 5G wireless HetNet is obviously an important issue to address. In fact, mobile operators would only optimize the placement and planning of certain network nodes (BSs), namely the numbers and locations of macro, micro, and pico BSs. This optimization must be performed considering that a large number of low-power femtocells are unplanned and installed quite arbitrarily by the end users. Moreover, the fact that large amount of mobile data traffic can be offloaded from one RAT to another tier of the same RAT or a different RAT must be taken into account in the network planning. This design task can be even more complicated since the network capacity offered by the indoor infrastructure may not be easy to estimate and can indeed vary over time.

Hence, existing planning tools for site locations such as using evenly spaced lattice structure with hexagonal grid is unlikely a good solution for all network tiers even though it may be still a reasonable solution for the macro BSs. In the current cellular network, the site locations and zoning of the BSs are determined from limited but expensive field test while the link and system level modelings take into account the carrier frequency, propagation environment and antenna characteristics. *However, the usefulness of this site acquisition approach may need to be verified for the 5G wireless HetNet.*

On the other hand, as the 5G network can be operated at both microwave and mmWave frequencies, the optimal frequency management and planning strategy is also not clear. This is especially because the cellular network has asymmetric uplink and downlink traffic demands. Furthermore, the mmWave frequencies have high penetration losses, which are easily absorbed or scattered by gases [14]. Consequently, the frequency management strategy at mmWave frequency bands may perhaps be location dependent (e.g., more mmWave frequency bands can be assigned when the surrounding environment has less buildings). *In summary, the network planning and optimization of the 5G network should take into account several unique parameters such as high cell density, extremely time-varying traffic and the fact that various frequency regions have quite different propagation characteristics.*

### B. Radio Resource Management

The capacity of future 5G networks would increase significantly with considerably higher cell densification, usage of higher mmWave frequencies and the advanced massive MIMO technologies. However, this heterogeneous network architecture poses several challenges for radio resource management. At mmWave frequency bands, the channel vector of each UE would have the specular characteristic which is the result of very few scatterers. Consequently, the beam pattern of a given UE is concentrated mainly around the direction of LOS path. Thus, UEs having different LOS paths would minimally interfere with each other. However, since the signal power of mmWave communication systems experiences high attenuation and the transmission bandwidth of mmWave systems is fairly large, noise will be a limiting factor for mmWave frequency communications. On the other hand, microwave channels are formed from several scatterers arising from different directions. Consequently, the beamforming pattern designed for one UE would create strong interference to all other UEs. For this reason, interference is a major limiting factor for microwave communication systems [6]. Thus, interference management is very crucial for microwave frequency bands. In particular, inter-user and inter-cell interference becomes more challenging to manage in the dense environment. This is because the cross-tier interference between macrocells and small-cells can only be mitigated efficiently if heavy coordination between the network tiers is performed by using high-speed and reliable backhauls, which are difficult to realize in practice.

Simple interference management methods based on orthogonal spectrum allocation approaches can be applied [2], which are, however, not so efficient for the dense 5G HetNet operating in the limited microwave spectrum. In fact, the Federal Communications Commission (FCC) in USA have found that more than 80% of the microwave frequency spectrum bands are not utilized efficiently. Thus, one way of improving the spectral efficiency of these bands is to adopt a more flexible management approach of available spectrum using the cognitive radio (CR) technology. In particular, the spectrum utilization of the 5G network can be improved by dynamically detecting the unused spectrum (i.e, white spaces) using different spectrum sensing algorithms. Then, interference-free transmissions between UEs and small-cell (macrocell) BSs can be enabled using the detected white spaces [18]. Different non-orthogonal spectrum management approaches have also been recently developed but most of them require cooperation among BSs of different tiers. *Thus, it is preferable that the future 5G network relies on autonomous interference management techniques that enable low-complexity, distributed solutions and possibly CR technologies.*

Furthermore, in recent years, 3D and FD MIMO beamforming techniques have been promoted to increase the spectral and energy efficiency of MIMO systems since these techniques would allow cellular systems to support a large number of UEs using multiuser MIMO techniques both for single cell and CoMP multicell systems. On the other hand, different requirements of the 5G network can be in conflict in such a way that improving one will degrade the others. To efficiently exploit all available resources, the future 5G network design must handle the existence of multiple objectives and inherent tradeoffs among them [19]. In particular, backhaul constrained multi-objective CoMP FD MIMO transceiver design presents an interesting design example with unique challenges. To this end, massive MIMO systems provide a large degrees of freedom, which not only considerably increases the system spectral and energy efficiency but also fundamentally eases the interference management. For instance, in a multicell setup, the massive MIMO capacity gains can be achieved with simple and uncoordinated beamforming schemes such as MRT or ZF (i.e., whenever there is no pilot contamination) [11]. However, since complete removal of pilot contamination is not a trivial task[2], CoMP transmission could be employed to reap the full benefits of massive MIMO systems. However, enabling CoMP transmission for massive MIMO systems may require significant coordination between different cells which may be practically infeasible as the number of UEs and antennas deployed in the system is likely very large. In contrast, certain degrees of freedom (i.e., antenna elements) offered by the massive MIMO can be leveraged to mitigate or cancel multi-cell and/or cross-tier interference via suitable beamforming techniques. *How to fully exploit massive MIMO to mitigate the interference among different cells with negligible BS coordination, and handle multi-objective designs is a non-trivial challenge.*

The low-power BSs of the 5G HetNet can be installed and managed by both wireless operators (such as pico BSs) and end users (such as femto BSs). Unlike the operator-installed low-power BSs, the low-power BSs which are deployed by the end users are installed quite arbitrarily. In addition, the unplanned low-power and indoor BSs can significantly outnumber the planned ones while the low-power BSs would be the main capacity drivers of the future 5G network since more than 70% of the traffic arises from the indoor environment. Therefore, these low-power nodes should possess self-configuration, self-optimization, and self-healing functionality (i.e., self organized network (SON)) in resource and interference management so that efficient and scalable ultra dense network deployment can be achieved. *In this regard, ensuring cost-effective customer satisfaction will be the critical challenge of the SONs. On the other hand, the resource allocation approach of the 5G network, unlike the traditional approach (which rely mainly on bit error rate), may need to consider different context information such as application, environment, and required QoS in terms of the latency, bit error rate, and minimum data rate requirements.* The context considers the user location, mobility, other proximity devices, resolution, central processing unit (CPU), battery level of the device. *How to deploy the SON that can enable a cost-effective context-aware resource allocation approach is a challenging problem.*

The 5G network is expected to achieve a minimum rate of 1Gb/s in an energy-efficient manner. One potential way to achieve these goals is to adopt the *virtual cell (soft cell)* concept which has been proposed for the dense wireless HetNet [20]. Specifically, radio resources available to a group of heterogeneous cells (macro, micro and femto cells) are considered a single resource pool to serve each UE, which in turn views the group of cells as a big macrocell. Furthermore, the idea of *phantom cell*, which was originally proposed by NTT DoCoMo, can be employed to better manage the control and data planes at different frequencies and nodes [21]. Specifically, the control and data planes are split so that critical control data is transmitted over reliable microwave links between UEs and macro BSs while high-speed data communications between UEs and small-cell BSs are realized over mmWave frequency bands. Consequently, a reliable and stable communications can be maintained while reaping the benefits of mmWave bands[3]. Also, macro BSs are utilized to ensure a wide-area coverage so as to maintain good connectivity and mobility. On the other hand, in a massive deployment of small cells, separating the control and data planes will improve the energy efficiency of the network [21]. *These design approaches advocate the user-centric principle where much more research studies are expected to realize it in practice.*

### C. Cell Association and Mobility Management

In the dense HetNet with multiple tiers and multiple RATs such as cellular technologies of different generations, WiFi and Worldwide Interoperability for Microwave Access (WiMaX), each UE can have several cell association options during the lifetime of its active sessions. Cell association should be designed to efficiently utilize the network resources and provide acceptable QoS for UEs. In general, small-cells such as picocells, femtocells, and WiFi are preferable to serve low mobility UEs whereas, high mobility UEs need to be served by the macro BS to avoid frequent handover. Furthermore, macro BSs should also fill the coverage holes of the network to reduce the call drop probability. For instance, in a HetNet network, a typical pico BS has a maximum cell radius of around 200m, and if one tries to associate a high-speed train UE traveling at 250 km/hr to this BS, handover may need to take place in few seconds, which is quite undesirable. Nevertheless, the development of an optimal cell association metric taking into account different factors such as signal quality, interference, traffic loads, data offloading capability and mobility is one of the key research issues. Moreover, the employed metric should enable decentralized implementation with low-signaling overhead. Traditional association metrics based on signal strength or signal to noise ratio would not be sufficient for the future 5G HetNet. Obviously, the performance of the 5G network can be improved further if each UE can be associated with more than one nodes as in the *virtual cell*. Moreover, cell association must be jointly designed with the mobility management so that low speed and high speed UEs can be treated differently. *In fact, high speed UEs are usually inside vehicles which should be served by macro BSs to avoid frequent handoffs. Furthermore, since vehicles can enable us to deploy massive MIMO, the reliability of these UEs can be improved significantly (see Fig. 3 in Section V).*

Offloading UEs to another tier of the same RAT or to another RAT offers an effective strategy to meet the requirements of the future 5G HetNet. In this respect, a fraction of user traffic can be rerouted either to another tier of the same RAT or to another RAT. Cell range expansion is one practical approach to offload traffic from macro BS to low power nodes. To further improve the offloading capability, operators can deploy their own WiFi APs to relieve the traffic congestion. These approaches may not be an ideal solution for the future network. To achieve the requirements of the 5G network, the offloading approach for the 5G network may need to exploit the advantages of all RATs. In a typical metropolitan environment where a number of collocated RATs exists, *discovering the best RAT for traffic offloading is not trivial*. This is because different RATs can be administered by different operators (in the case of WiFi AP for example). On the other hand, as inter-RAT offloading utilizes the resources of other RAT, *the offloading revenue should be shared between the RAT operators fairly*. This offloading approach is particularly useful to *disseminate non-private and delay tolerant* information such as bulk data files, news, software and files generated by laboratory experiments, and sensor networks which account for 64% of the current world mobile data traffic. Furthermore, the offloading method may need to consider location, user, and time-dependent RAT availabilities, incorporate both licensed and unlicensed bands at mmWave frequencies such as WiGIG, and take into account the seamless service delivery switching time constraint between different RATs which is around 10ms [1]. *Designing a cell association and traffic offloading algorithm by integrating all these issues is a non-trivial and challenging problem.*

---

[2]Note that one can efficiently mitigate the effect of pilot contamination (see Section IV-E).

[3]Sporadic interruptions of the data plane during the handover process between two small cells become less harmful.

### D. Backhaul Design

Backhaul links are required for data and signaling exchanges between BSs and between the core and access networks (i.e., the system of BSs). In general, high-capacity and reliable backhaul links support different types of traffic and cooperation between different BSs, which consequently improves the user experience and overall throughput of the network. There has been active research on the advanced CoMP transmission and reception techniques for the LTE-based system over the past few years. For the future dense HetNet, deployment of an efficient backhaul network supporting coordination and signaling among BSs of different types is undoubtedly required. Studies in 2012 show that the backhaul of the existing cellular network comprises of 70% cooper, 10% fiber and 15% wireless backhaul. And it would be preferable to deploy more wireless over wired backhaul technologies although wired backhaul using optical fiber and internet protocol (IP) cables would still be needed in the 5G network. In the existing wireless backhaul, large physical aperture antennas are used to achieve the required link gain which is not economically desirable. *Therefore, engineering of the cost-effective, reliable, and scalable wireless backhaul network will be one of the fundamental challenges of the future 5G network.*

MmWave and massive MIMO technologies provide great opportunities to resolve these challenges. Specifically, one can form a large number of beams to establish point-to-multipoint backhaul links by using massive MIMO [11]. In addition, mmWave offers virtually unlimited bandwidth for short-range backhaul links, which would be sufficient for the future dense HetNet (For instance, around 12.9 GHz of bandwidth is available in the E-band). This is illustrated in the hierarchical (mesh) backhaul network shown in Fig. 1. The massive MIMO beamforming, if designed appropriately, can realize mmWave narrow beams which provide a unique opportunity to have scalable backhaul network with minimal (negligible) interference. One can utilize simple resource allocation approaches such as static partitioning (usually in frequency domain) for the access and backhaul networks. A more efficient approach is to allow the *backhaul and access networks share the same resource pool*. By doing so, the deployment flexibility of the network can be improved [15]. In this direction, some research works are performed including design, resource allocation and scheduling mechanisms [22]. However, the unique challenge of the latter approach is that as a low-power node operates on the same frequency both for backhaul and access links, each node may need to allocate backhaul resources dynamically according to its load and the meshed low power nodes. *For this reason, re-engineering the backhaul may need to be considered taking into account the time varying load characteristics of all low-power nodes.*

### E. Low-Cost CSI Acquisition and Beamforming for Massive MIMO

Beamforming is the key tool to exploit the potentials of MIMO systems, which typically requires the availability of channel state information (CSI) at the transmitter and/or receiver. For the future 5G network that employs massive MIMO at the microwave and/or mmWave frequency bands, CSI estimation and beamforming are the key design issues which strongly impact the network performance. The existing MIMO systems (such as LTE) are equipped with a few number of antennas $N$ (from 1 to 10). For such a case, the number of radio frequency (RF) chains, digital to analog converters (DACs) and analog to digital converters (ADCs), which are the most expensive parts of a wireless transceiver, can be the same as that of the number of antennas. However, in a massive MIMO system with 100 to 1000 antennas, deploying $N$ RF chains is not practically feasible[4]. On the other hand, the energy consumptions of a MIMO transceiver increases with the number of active RF chains. Specifically, when the transmission bandwidth is very large, the energy consumption of ADCs would be unacceptably high. *Thus, the channel estimation and beamforming algorithms should be designed taking into account the constraints on the number of RF chains (i.e., to be much less than $N$) and the finite resolutions of ADCs.*

In a conventional channel estimation (i.e., $N$ RF chains case), the channels between transmitters and receivers are estimated from orthogonal pilot sequences which are limited in number due to the finite coherence time of the channel. In a massive MIMO setup, as $N \gg K$ (i.e., the number of UEs), time division duplex (TDD) based communication would be more efficient where the pilot symbols are transmitted from UEs [11]. In a multicell setup, however, orthogonal pilot sequences may need to be re-used over the cells which unfortunately results in the so-called *pilot contamination*. Pilot contamination imposes the fundamental performance bottleneck for the massive MIMO system[5]. Therefore, research on potential techniques to eliminate or mitigate the pilot contamination effect through efficient CSI estimation and pilot optimization is important [11]. Different approaches have been proposed to reduce pilot contamination such as Eigenvalue decomposition (EVD) based channel estimation, successive pilot transmission, time shifted channel estimation and a linear combination approach which exploits multipath components [23], [24]. In this regard, it is shown recently in [24] that a maximum of $L$ (i.e., number of multipath components) cells can reliably estimate the channels of their UEs and perform beamforming while efficiently mitigating the effect of pilot contamination where $L$ can be increased just by increasing the transmission bandwidth. Nevertheless, all these approaches can only mitigate the effect of pilot contamination and, how to completely eliminate the effect of pilot contamination is still an open research problem.

---
[4]Note that the number of RF chains is the same as that of DACs at the transmitter, and ADCs at the receiver. Thus, we simply refer to RF chains as they implicitly include, ADCs and DACs in the sequel.

When the number of RF chains $N_{RF}$ at the BS is much less than the number of BS antennas, the orthogonal channel estimation approach cannot be used even for single cell setup (i.e., without pilot contamination). Towards this end, there are two potential methods. The first method is to employ a combined *time-frequency analog-digital channel estimation approach*. In such approach, the high dimensional section of the channel estimation process will be commonly employed for all UEs and sub-carriers in time domain and analog form (i.e., depending on $N$). And the low dimensional part (i.e., depends on $N_{RF}$) can be designed in digital form for each UE and sub-carrier. In general, this approach is effective for an environment where the channel parameters have several multipath components (e.g., microwave frequency bands). The second method is to leverage the idea of *direction of arrival (DOA) estimation* approach from radar signal processing to mmWave channel estimation. The mmWave frequency bands are effective in a LOS environment and detecting the DOA of the transmitted signal is sufficient in such an environment. In this regard, one can estimate the DOA at microwave frequencies and employ this DOA information for data transmission at mmWave bands. *Specifically, this approach is quite interesting when the low-power node is capable of operating at both mmWave and microwave frequency bands such as a WiGIG access point*[6].

Once the channel estimation process has been completed, the next step is to perform beamforming. Again, the conventional MIMO system is based on digital beamforming which is performed at the baseband level and it requires $N_{RF} = N$, which is very costly. To reduce the *implementation cost* of beamforming, a combined *time-frequency analog-digital approach* is one of the most promising ones. This beamforming approach will have both high dimensional analog part operating in the time domain, and low dimensional digital part designed for each sub-carrier [13]. It is well-known that the channel estimation and beamforming are interrelated; when the channel estimation is more inaccurate, the performance of the beamformer can become worse. *Therefore, the critical challenge is how to jointly and optimally perform channel estimation and beamforming taking into account the number of RF chains and pilot contamination for the multiuser multi-cell massive MIMO system operating over the frequency selective channel.*

## V. CASE STUDIES

In the following, we discuss some design and case studies, which illustrate how one can address some challenges of the 5G MIMO HetNet presented in the previous section.

### A. Low-Cost Hybrid Digital-Analog Beamforming for Multiuser Massive MIMO

We demonstrate the desirable cost and performance tradeoff achieved by the hybrid beamforming for the multiuser massive MIMO system with large-scale antennas at both the transmitter and receiver sides [13]. As detailed earlier, the key advantages of hybrid beamforming is that it requires smaller number of RF chains, ADCs and DACs compared to the number of transmitter and receiver antennas, respectively. To investigate the performance achieved by hybrid and digital beamformings, we adopt a geometric channel model with $L$ scatterers, where each scatterer is assumed to contribute a single propagation path between the BS and each UE. Furthermore, the BS and each UE are equipped with uniform linear arrays (ULA) antenna with half wavelength spacing. The number of transmitter antennas, transmitter RF chains, receiver antennas and receiver RF chains are set equal to 128, 64, 32 and 16, respectively. The number of receivers is $K = 4$, the number of multiplexed symbols for each receiver is $S_k = 8$ and $L = 16$. Fig. 2 shows the sum rate of the digital and hybrid beamformings for the downlink multiuser massive MIMO system. *As can be seen from this figure, the hybrid beamforming approach of [13] achieves almost the same performance as that of the digital beamforming.*

The result in this figure is obtained for the channel parameters discussed in this subsection. Thus, one could wonder whether the hybrid design of [13] can achieve the same performance as that of the digital one for arbitrary channel parameters. In this respect, it is demonstrated via simulation in [25] that the hybrid beamforming design of [13] can experience significant performance loss (compared to digital beamforming) when the number of scatterers are large. This motivates [25] to develop a hybrid beamforming utilizing a fixed (digitally controllable) paired phase shifters and switches, that achieves closer performance to that of the fully digital digital beamforming. This paper also determines the required number of RF chains (and phase shifters) such that the hybrid design achieves the same beamforming performance as that of the digital one. Moreover, for the parameter setups of this subsection, it is shown that the hybrid design requires a maximum of $KS_k$ and $S_k$ RF chains at the transmitter and receiver, respectively, to maintain the same performance as that of the digital design [25].

---

[5]In a traditional multicell multiuser MIMO system, although pilot contamination arises, its effect is negligible or can be alleviated by employing appropriate coordinated BS transmission and CSI acquisition approaches.

[6]Note that in areas where there are multiple reflected rays, the DOA information of mmWave and microwave frequency bands may be different from each other. For such scenario, environmental (context) aware DOA estimation approach can be employed jointly with the CR technology [18].

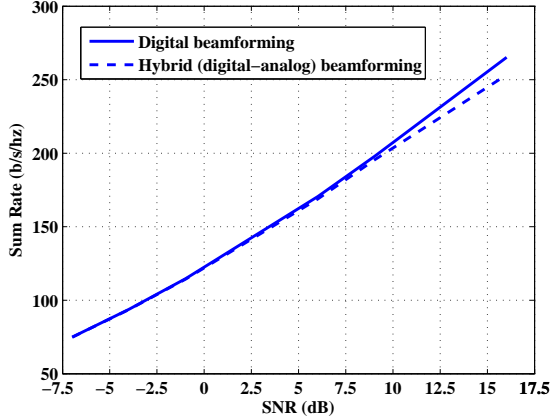

Fig. 2. Comparison of digital and hybrid beamformings.

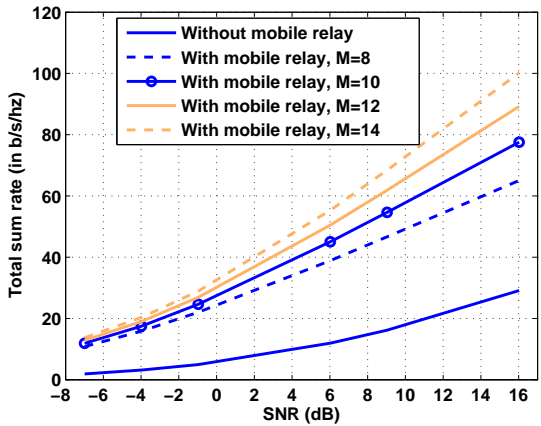

Fig. 3. Comparison of the sum rates achieved with and without mobile relay. In this figure, $M$ denotes the number of antennas at the mobile relay.

### B. Mobile Relay with Large-Scale Antenna Array

We demonstrate the performance gain achieved by implementing large-scale antenna arrays at mobile relays which can be deployed on public transportation vehicles such as trains or buses as depicted in Fig. 1 to better support highly mobile UEs. In particular, we consider two downlink scenarios where a macro BS communicates with individual UEs directly and through the mobile relay which are refereed as first and second scenarios, respectively. In the second scenario, the communication between UEs inside the vehicle and mobile relay are performed at mmWave frequency bands where there is almost no capacity limitation. The BS is located at the center of its cell whereas, each UE is located randomly inside the vehicle of dimension 3.10m × 25.91m (Amtrak passenger train) with uniform distribution. The propagation model between the BS and each UE contains two components.

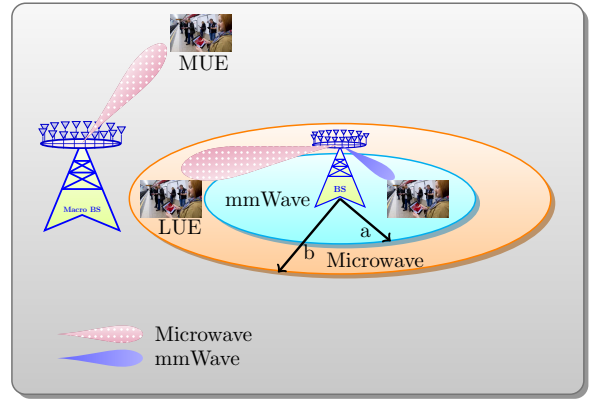

Fig. 4. Suggested dual band, layered and prioritized cell association. MUE (LUE) denotes UE served by the macro BS (low power node).

One is the path loss component decaying with distance, and the other one is the Rayleigh fading random component which has a zero mean and unit variance. All the other parameter settings are summarized as shown in Table I.

TABLE I
SIMULATION PARAMETERS FOR FIG. 3.

| Distance between macro BS and Vehicle | 1km |
|---|---|
| Number of antennas at each UE | 2 |
| Macro BS transmitter power | 5W |
| Radius of the cell | 1.6km |
| Reference distance ($d_0$) | 1.6km |
| Path loss exponent | 3.8 |
| Mean path loss at $d_0$ | 134dB |
| Channel bandwidth | 5MHz |
| Carrier frequency | 1.8GHz |
| Receiver noise figure | 5dB |
| Receiver vertical antenna gain | 10.3dBi |
| Receiver temperature | 300K |

We compare the total rates achieved in these two scenarios. For the first scenario, we consider that the macro BS employs multiuser block diagonalization (BD) beamforming approach. In the second scenario, since the mmWave frequency band transmission is assumed to support very high rate, the achievable rate is determined by the radio link between the macro BS and mobile relay. For this scenario, the macro BS employs singular value decomposition (SVD) beamforming. Fig. 3 shows the sum rate computed by the Shannons capacity formula of these two scenarios. As we can see from this figure, the rate achieved by the second scenario (i.e., with mobile relay) is significantly higher than that of the first scenario (i.e., without mobile relay). In addition, we also see that increasing the number of mobile relay antennas improves the achievable rate which is indeed expected. *These results confirm that a large-scale antenna system (massive MIMO) can lead to the significant performance gain when utilized appropriately.*

### C. Dual-Band Small-Cells

We now discuss a potential design for a small-cell network that operates in both microwave frequencies and mmWave frequencies [15]. In general, microwave frequencies have more favorable propagation properties but are more limited in bandwidth compared to the mmWave frequencies. *Thus, a dual band small-cell with prioritized and layered cell association strategy as shown in Fig. 4 can exploit the advantages of these different frequency bands.* In this design, the coverage area is divided into three regions where the UEs in the inner region (i.e., radius a)/middle region (i.e., radius b) are served by the small cell at mmWave/microwave frequency bands. And the UEs in the outer region (i.e., radius $> b$) are served by the macrocell at microwave frequency bands. Here, the outer region operates at microwave frequency bands since these bands allow long-range communications to support mobility. The parameters $a$ and $b$ can be chosen or optimized adaptively based on different *context information* as discussed in the previous section. The middle region (i.e., with radius b) is analogous to the "expanded range" considered in the LTE standard where its radius can vary from one design criteria to another [10]. *This dual-band small-cell design isolates the mmWave and microwave frequency UEs (i.e., the mmWave UEs inside the radius $a$ do not experience any interference from the macro BS which is desirable). However, still how to choose the best radius for different regions again depends on specific design objectives, which present interesting open research problems.*

## VI. CONCLUSIONS

This article discusses the 5G wireless HetNet integrating massive MIMO and mmWave technologies. We have also presented a potential HetNet architecture that employs both microwave and mmWave frequencies. Various design and technical challenges for this network architecture have been described. We have then presented three case studies addressing some of the challenges and showing their benefits. Overall, this article lays out the potential architecture and points out the underlying research challenges, which must be addressed to meet the technical requirements of the future 5G network.